# Water quality information dissemination at real-time in South Africa using language modelling

**Laing Lourens***
Next Generation Technologies
Meraka Institute, CSIR
Pretoria, South Africa
lainglourens@gmail.com

**Arijit Patra***
University of Oxford
Oxford, United Kingdom
arijit.patra@eng.ox.ac.uk

**Luqmaan Hassim***
ICT4EO
Meraka Institute, CSIR
Pretoria, South Africa
luqmaanhassim1@gmail.com

**Faheem Sima**
SATSAR
Meraka Institute, CSIR
Pretoria, South Africa
Faheem.sima7@gmail.com

**Avashlin Moodley**
Human Language Technologies
Meraka Institute, CSIR
Pretoria, South Africa
avashlin@gmail.com

**Pulkit Sharma**
University of Oxford
Oxford, United Kingdom
pulkit.sharma@eng.ox.ac.uk

## Abstract

We present a conversational model to apprise users with limited access to computational resources about water quality and real-time accessibility for a given location. We used natural language understanding through neural embedding driven approaches. This was integrated with a chatbot interface to accept user queries and decide on action output based on entity recognition from such input query and online information from standard databases and governmental and non-governmental resources. We present results of attempts made for some South African use cases, and demonstrate utility for information search and dissemination at a local level.

## 1 Introduction

Developing countries have non-trivial constraints when it comes to adequately monitoring local environments and disseminating relevant information around them. With the advent of smartphones and cheap data plans, it is pertinent to alleviate such information asymmetry by leveraging the wealth of information available in online content generated across sources and estimate the locally relevant pieces using machine learning based tools reliant on natural language process and similar methods. The idea was to incorporate such automated tools to extract topics of situational relevance for different stakeholders in the constraints of inadequate computational bandwidth, information distribution at the last mile and a lack of physical infrastructure or capital resources. Such a topical exploration of relevant information using a machine learning driven interactive chatbot system has myriad applications in our action areas, like locating potable water sources with real-time information about impediments to accessibility owing to unanticipated events such as road closures or unrest. Such a pipeline can be also used for other use cases like cyclone warnings, disease outbreaks etc. Recent advances in automation of pattern recognition processes in unstructured data has been largely influenced by the emergence of deep learning methods in image understanding (1; 2) and language (3; 4; 5). We attempt to leverage the progress made on applications around deep learning for language, speech (6) and image context understanding to explore solutions towards water quality information modelling and similar environmental aspects.

---

*equal contribution

We consider the problem of accessibility of water resources with considerations to the effects of real-time impediments driven by unforeseen situations. In this work, we attempted to implement the information relay system using a combination of a Natural Language Understanding unit and a dialogue module implemented as a chatbot which serves as the user input and the pipeline output interface recommending a policy/action based on the attentive information from the user extracting keywords or key phrases (such as water quality) and location along with relevant metadata, and matching such requests with relevant sources obtained from online mining of concerned databases/websites to provide query-relevant knowledge. We leverage the recent works on entity recognition to estimate information relevance and bundled it with attention mechanisms as described in Fig. 2. The implementation in the South African context is at a community level and the idea has been to improve access to information for local, granular decision making. The utility of the language processing modules has helped in parsing relevant governmental, non-governmental and media resources to address information gaps in such decision making.

## 2   Model

Our proposed architecture is based on a NLP component, and an interface in the form of a chatbot (detailed in figure 1). The user input query is processed using a Natural language understanding model based on the RASA toolkit (5) for processing sequences in English (to be extended to other South African languages) to classify and parse semantically and contextually relevant information. The NLU model was trained with the tensorflow embedding pipeline. The policy for this pipeline has multiple steps, including: using dense layers to establish embeddings for entities and actions, including their histories; using Neural Turing Machines (NTM) for calculating attention probabilities over system memory (3); detailed use of LSTM to determine similarity between dialogue embedding and embedded system actions (4). This allows us to perform entity recognition and intent modelling following the language identification step. A recurrent attention mechanism on top of this is used to identify salient keywords to aid with the global search from publicly available information with respect to water quality at real time.

Real-time availability of water and its accessibility is typically a function of complex interplays of predictable data and unpredictable stochastic impediments, which can not be estimated from static data alone. Therefore, it is important that our machine learning based method is able to incorporate real-time changes in the information state at the input stage. We identify key variables with respect to the last mile access to potable water resources not only in terms of environmental variables, but also social and situational aspects like outbreaks of epidemics, civil unrest etc. We term this dictionary as the 'Situational Variables' used to inform the Action step for the chatbot response once the policy has been decided to be apart from Greeting or Farewell (and based on the conversation state which is tracked for the bot). This is so that such stochastic events be tracked from user input and publicly scraped information from sources like the Department of Water Affairs (Govt. of RSA), WESSA (Wildlife and Environmental Society of South Africa) and Cyanolakes (online monitoring and mapping service for water and health authorities). Although API's are not yet available for all these data sources, they are included as such in the proposed architecture for optimal system functionality. Instead, subsets of these data sources were stored locally for testing purposes. These sources are parsed to inform the Action of the chatbot for response to the original user queries.

Such determinants are fused with the extracted salient representations from user input query to construct the environmentally aware real-time information generator. Thus, the overall pipeline is a combination of keyword based methods and sequence modelling. The motivation behind such fusion is to enable effective resilience to noisy sources of user input and a non-deterministic diversity of impediments encountered in real use cases with regards to directing communities to accessible water sources. A direct implementation of a natural language understanding model to process requests at this level comes with a high probability of providing static information which may be too generalist and temporally out of sync with actual accessibility states for a given location. Also, the modules are trained to detect situation keywords in user inputs, so as to expand the keywords database with multiple local instances of co-occurrences with water resource data. This enables the training of our model to be pursued in an online fashion with user feedback.

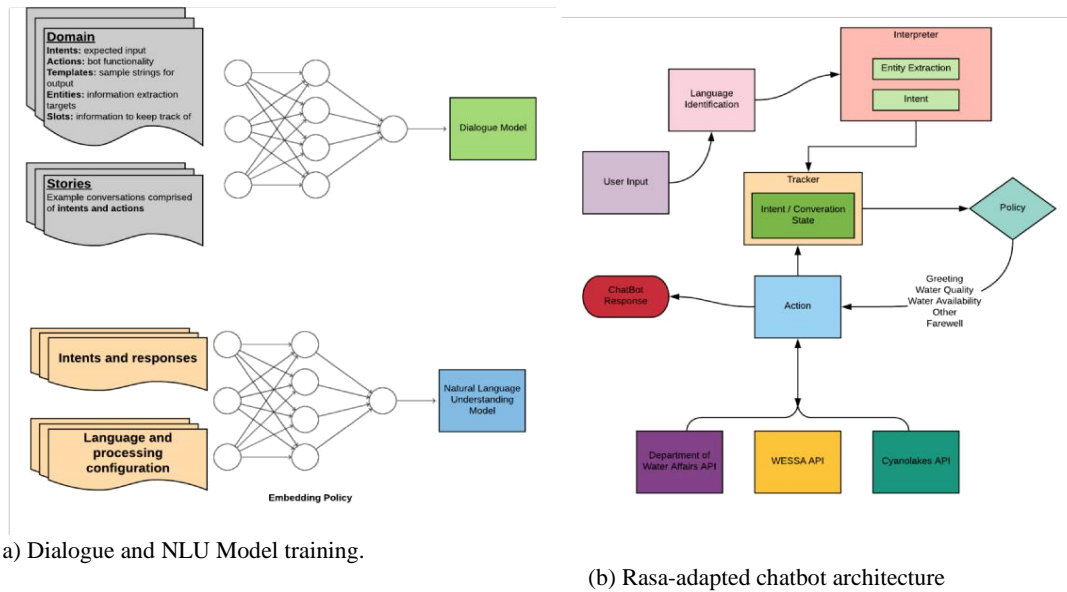

a) Dialogue and NLU Model training.

(b) Rasa-adapted chatbot architecture

Figure 1: NLU and Chatbot overview

## 3 Results

We attempted the proposed system with respect to the case of searching for water quality information about different areas in Cape Town, with respect to both potable water, restrictions on usage and coastal water quality and safety. The confusion matrices demonstrate the ability to categorize the query and arrive at the correct conversation state. The precision, recall and F1 scores are used to quantify the performance over different classes of utterances possible.

(a) Example response for a safe drinking supply.

(b) Example response for a drinking supply that is not safe.

Figure 2: Information Dissemination Examples

### 3.1 Conversational Authenticity and Interactive Learning

The initial implementation revealed there were an inadequate number of stories provided for the model in the training data. For instance, the chatbot would bid farewell when asked a question. These issues were mostly resolved once stories that allowed multiple questions after (or without) a greeting. As can be seen in figure 2, adequate results were demonstrated with a simplified water quality model, where Cape Town has water that is safe to drink and fictional Escape Town does not have safe water. This demonstrates that the model was able to identify the user's intent, store the location in its state and lookup relevant information based on the data provided as input.

Figure 3: Interactive Learning example

The Rasa framework provides an interactive learning mode, where live feedback is provided to the bot figure 3. This has the benefit of guiding the chatbot's communication patterns and is a simple but effective method of correcting mistakes. The output of this interaction was combined with the original training data and the model was retrained with this augmented data set.

### 3.2 Model testing and improvement

Following the data augmentation, the average recall of the updated model increased from 83% to 85%, while the average f1-score increase from 85% to 87%. The average precision of the augmented model dropped from 93% to 92%. From the confusion matrix we can see that the beach quality utterance was the response most commonly misclassified by the model. The complete output of model performance can be seen in table 1 and a visualisation of the confusion matrix can be seen in figure 4. Identical results were observed when training the NLU model with the spacy scikit learn pipeline and the embedding method.

Table 1: Model performance

|  | Precision | Recall | f1-score | Support |
| --- | --- | --- | --- | --- |
| action_default_fallback | 0.00 | 0.00 | 0.00 | 0 |
| action_listen | 1.00 | 1.00 | 1.00 | 118 |
| utter_beach_quality | 0.80 | 0.48 | 0.60 | 25 |
| utter_goodbye | 0.78 | 1.00 | 0.88 | 18 |
| utter_greet | 0.74 | 0.91 | 0.82 | 34 |
| utter_water_availability | 1.00 | 0.55 | 0.71 | 22 |
| utter_water_quality | 1.00 | 0.47 | 0.64 | 19 |
| Average / Total | 0.92 | 0.85 | 0.87 | 236 |

## 4 Future Work and Discussion

We proposed and validated a information dissemination model integrating a machine learning driven language modelling task with an interactive chatbot infrastructure. The utility was shown for different applications pertinent to water quality in different circumstances and locations, in a user-friendly conversational style. Possible future extensions would look at integrating insights from sentiment analysis of social media mentions, refining the natural language embedding process using seq2seq models and modelling conceptual level information, apart from expanding to different languages and use cases in Southern Africa and beyond. A direction of research also needs to be explored on the mechanisms for secondary and tertiary information ingestion into the system, such as building and identifying API's for relevant data. It is noted that the overall architecture is lightweight and has a low computational requirement, and is amenable to deployment on low cost smartphones.

## 5 Acknowledgements

Arijit Patra is supported by The Rhodes Trust.

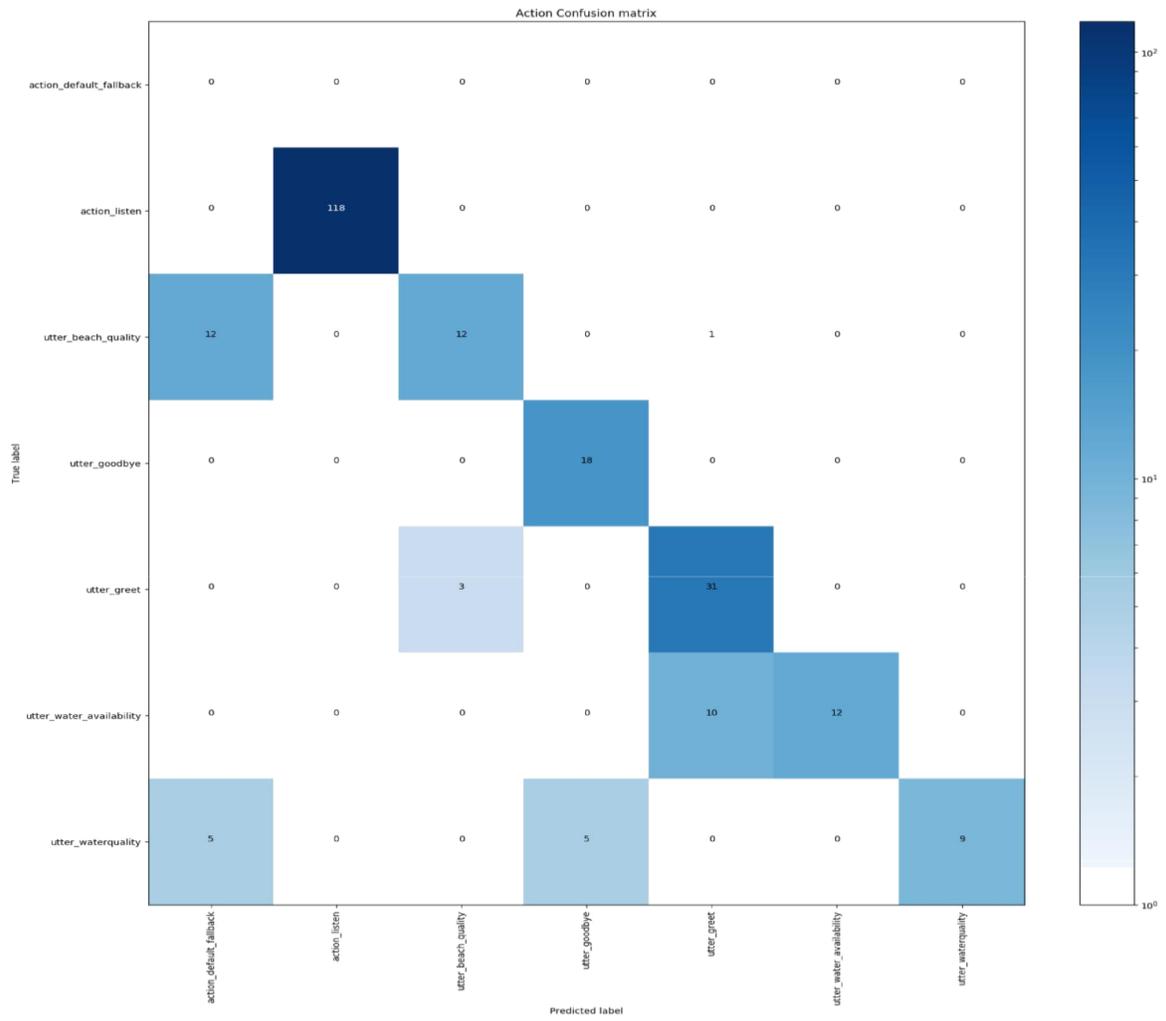

Figure 4: Action Confusion Matrix